\documentclass[preprintnumbers]{revtex4}
\UseRawInputEncoding
\usepackage{amssymb}
\usepackage{amsmath}
\usepackage{graphicx}
\usepackage{dcolumn}
\usepackage{bm}
\usepackage{subfigure}
\usepackage{color}

\begin{document}

\title{Nonlinearity effect on Joule-Thomson expansion of Einstein-power-Yang-Mills AdS black hole}
\author{Yun-Zhi Du$^1$, Xiao-Yang Liu$^1$, Yang Zhang$^1$, Li Zhao$^2$\footnote{the corresponding author}, and Qiang Gu$^3$}

\address{$^1$Institute of Theoretical Physics, Shanxi Datong University, Datong,  China\\
$^2$ Lanzhou Center for Theoretical Physics and Research Center of Gravitation, Lanzhou University, Lanzhou, China\\
$^3$ Department of Basic Construction, Shanxi Datong University, China}

\thanks{\emph{e-mail:duyzh22@sxdtdx.edu.cn, $sxdtdx\_liuxy@163.com$, zhangyphysics@126.com, lizhao@lzu.edu.cn, gudujianghu23@163.com}}

\begin{abstract}
Considering the nonlinearity of the Yang Mills charge, we investigate the Joule-Thomson expansion for the Einstein-Power-Yang-Mills AdS black holes in the context of the gauge-gravity duality. Under this framework, we calculate the Joule-Thomson coefficient, describe all relevant inversion and isenthalpic curves in the temperature-pressure plane that determining in this manner the corresponding cooling and heating regions. Finally we analyze the effect of the charge nonlinearity on the Joule-Thomson expansion.
\end{abstract}

\maketitle

\section{Introduction}

In recent decades, people confirm the fact that black holes are thermodynamics systems \cite{Bardeen1973,Jacobson1995,Padmanabhan2009} where its area is related with entropy and its surface gravity is connected with temperature \cite{Bekenstein1973a,Hawking1975}. The following important step is to establish the quantum gravity theory. The negative cosmological constant in an Anti-de Sitter (AdS) spacetime with black hole will lead to phase transitions of black holes \cite{Hawking1983,Witten1998}. And the corresponding conjugate quantity of pressure in an extended phase space for black holes is the volume \cite{Kastor2009}. The physical implication was related with the holography, where black holes would being a system and dual to conformal field theories \cite{Maldacena1998}. That makes AdS black holes can be identical to the thermodynamics of ordinary systems and their thermodynamics become more complete. Especially there exist several different types of phase transition, the Van de Walls (VdW's)-like phase transition \cite{Kubiznak2012,Cai2013,Wei2015}, reentrant phase transitions \cite{Altamirano2013,Frassin2014}, the polymer-like phase transition \cite{Dolan2014}, and the triple points \cite{Wei2014,Li2022}, along with the novel dual relation of HP phase transition \cite{Wei2020}. Meanwhile, the inclusion of the pressure-volume term in the thermodynamical first law makes other model parameters as novel thermodynamical quantities \cite{Cai2013} and make it possible to regard AdS black holes as heat engines \cite{Johnsom2014,Xu2017}. All of those developments are in the subdiscipline, black hole chemistry \cite{Kubiznak2017}.

In the classical thermodynamics, there is a well known process named the Joule-Thomson expansion, i.e., the gas moves from a region of high pressure to a region of low pressure via an equal velocity. Based on this, the JT effect of the charged AdS black hole was firstly investigated in ref. \cite{Okcu2017}. Subsequently the JT expansion becomes an active issue and gets more attention, furthermore is extended to the study of other black holes \cite{Okcu2018,Mo2018,Mo2020,Xing2021,Liang2021,Ditta2022}.
Additionally, at the linear level the charged black holes in an AdS spacetime nearby the critical point is of the scaling symmetries, $S\sim q^2,~P\sim q^{-2},~T\sim q^{-1}$ \cite{Johnson2018,Johnson2018a}. It is naturel to guess that whether the same scaling symmetry still hole on for the non-linear charged AdS black holes? There are lots of the generalization of the linear charged AdS black hole solution: Einstein-Maxwell-Yang-Mills AdS black hole \cite{Mazharimousavi2008}, Einstein-Power-Yang-Mills AdS black hole \cite{Lorenci2002}, Einstein-Maxwell-Power-Yang-Mills AdS black hole \cite{Zhang2015a}, Einstein-Yang-Mills-Gauss-Bonnet black hole \cite{Mazharimousavi2007}, Einstein-power-Maxwell-power-Yang-Mills-dilaton \cite{Stetsko2021}, and so on. An interesting non-linear generalization of charged black holes involves a Yang-Mill field exponentially coupled to Einstein gravity (i.e., Einstein-Power-Yang-Mills gravity theory) because it possesses the conformal invariance and is easy to construct the analogues of the four-dimensional Reissner-Nordstr\"{o}m black hole solutions in higher dimensions. Additionally several thermodynamical features of the EPYM AdS black hole in the extended phase space have been exhibited \cite{Zhang2015a,Du2021,Moumni2018}. Here we pay attention on the JT expansion for the non-linear charged AdS black hole in this theory.

In this paper we study and discuss the Joule-Thomson expansion for black holes in the model of nonlinear electrodynamics (NED) coupled to gravity in AdS spacetime. The interest to NED model \cite{Bronnikov2001} considered is due to its simplicity: the metric function is expressed via simple elementary function. This model was explored to study the supermassive black hole M87* \cite{Bronnikov2001} and to construct non-singular model of magnetized black hole \cite{Allahyari2020}. In Sec. \ref{scheme2}, we briefly review the EPYM AdS black hole solution and its hawking temperature. In Sec. \ref{scheme3}, we investigate the Joule-Thomson expansion for the EPYM AdS black hole. A brief summary is given in Sec. \ref{scheme4}.

\section{EPYM AdS black hole and Hawking temperature}
\label{scheme2}
The action for four-dimensional Einstein-power-Yang-Mills (EPYM) gravity with a cosmological constant $\Lambda$ was given by \cite{Zhang2015,Corda2011,Mazharimousavi2009,Lorenci2002}
\begin{eqnarray}
I=\frac{1}{2}\int d^4x\sqrt{g}
\left(R-2\Lambda-\mathcal{F}^\gamma\right)
\end{eqnarray}
with the Yang-Mills (YM) invariant $\mathcal{F}$ and the YM field $F_{\mu \nu}^{(a)}$
\begin{eqnarray}
\mathcal{F}&=&\operatorname{Tr}(F^{(a)}_{{\mu\nu}}F^{{(a)\mu\nu}}),\\
F_{\mu \nu}^{(a)}&=&\partial_{\mu} A_{\nu}^{(a)}-\partial_{\nu} A_{\mu}^{(a)}+\frac{1}{2 \xi} C_{(b)(c)}^{(a)} A_{\mu}^{(b)} A_{\nu}^{(c)}.
\end{eqnarray}
Here, $\operatorname{Tr}(F^{(a)}_{\mu\nu}F^{(a)\mu\nu})
=\sum^3_{a=1}F^{(a)}_{\mu\nu}F^{(a)\mu\nu}$, $R$ and $\gamma$ are the scalar curvature and a positive real parameter, respectively; $C_{(b)(c)}^{(a)}$ represents the structure constants of three-parameter Lie group $G$; $\xi$ is the coupling constant; and $A_{\mu}^{(a)}$ represents the $SO(3)$ gauge group Yang-Mills (YM) potentials defining by the Wu-Yang (WY) ansatz \cite{Balakin2016}. Variation of the action with respect to the spacetime metric $g_{\mu\nu}$ yields the field equations
\begin{eqnarray}
&&G^{\mu}{ }_{\nu}+\Lambda\delta^{\mu}{ }_{\nu}=T^{\mu}{ }_{\nu},\\
&&T^{\mu}{ }_{\nu}=-\frac{1}{2}\left(\delta^{\mu}{ }_{\nu} \mathcal{F}^{\gamma}-4 \gamma \operatorname{Tr}\left(F_{\nu \lambda}^{(a)} F^{(a) \mu \lambda}\right) \mathcal{F}^{\gamma-1}\right).
\end{eqnarray}
Variation with respect to the 1-form YM gauge potentials $A_{\mu}^{(a)}$ and implement the traceless yields the 2-forms YM equations
\begin{equation}
\mathbf{d}\left({ }^{\star} \mathbf{F}^{(a)} \mathcal{F}^{\gamma-1}\right)+\frac{1}{\xi} C_{(b)(c)}^{(a)} \mathcal{F}^{\gamma-1} \mathbf{A}^{(b)} \wedge^{\star} \mathbf{F}^{(c)}=0,
\end{equation}
where $\mathbf{F}^{(a)}=\frac{1}{2}F_{\mu \nu}^{(a)}dx^\mu\wedge dx^\nu,~\mathbf{A}^{(b)}=A_{\mu }^{(b)}\wedge dx^\mu$, and ${ }^{\star}$ stands for the duality. It is obviously that for the case of $\gamma=1$ the EPYM theory reduces to the standard Einstein-Yang-Mills (EYM) theory \cite{Mazharimousavi2007}. In this work our issue is paid on the role of the non-linear YM charge parameter $\gamma$.

Here we should point out that the non-Abelian property of the YM gauge field is expressed with its YM potentials
\begin{eqnarray}
\mathbf{A}^{(b)}=\frac{q}{r^2}C^{(a)}_{(i)(j)}x^idx^j,~r^2=\sum_{j=1}^3x_j^2,
\end{eqnarray}
and $q$ is the YM charge, the indices ($a,~i,~j$) run the following ranges: $1\leq a,~i,~j\leq3$. The coordinates $x_i$ take the following forms: $x_1=r \cos\phi \sin\theta,~x_2=r \sin\phi \sin\theta,~x_3=r \cos\theta.$ Since we have utilized the WY ansatz for the YM field, the invariant for this field takes the form \cite{Stetsko2020,Chakhchi2022}
\begin{eqnarray}
\operatorname{Tr}(F^{(a)}_{{\mu\nu}}F^{{(a)\mu\nu}})=\frac{q^2}{r^4}.
\end{eqnarray}
This form leads to the disappearance of the structure constants which can be described the non-Abelian property of the YM gauge field. Therefore, under the condition of the WY ansatz we may focus on the role of the non-linear YM charge parameter, instead of the non-Abelian character parameter.

The metric for the four-dimensional EPYM AdS black hole is given as follows \cite{Yerra2018},
\begin{eqnarray}
d s^{2}=-f(r) d t^{2}+f^{-1} d r^{2}+r^{2} d \Omega_{2}^{2},
\end{eqnarray}
where
\begin{eqnarray}
f(r)=1-\frac{2\bar M}{r}+\frac{r^{2}}{l^2}+\frac{\left(2q^{2}\right)^{\gamma}}{2(4 \gamma-3) r^{4 \gamma-2}}. \label{f}
\end{eqnarray}
Here $d\Omega_{2}^{2}$ is the metric on unit $2$-sphere with volume $4\pi$ and $q$ is the YM charge, $l$ is related to the cosmological constant: $l^2=-\frac{3}{\Lambda}$, $\gamma$ is the non-linear YM charge parameter and satisfies $\gamma>0$ \cite{Corda2011}. The event horizon of the black hole is obtained from the relation $f(r_+)=0$. The mass parameter of the black hole can be expressed in terms of the horizon radius as
\begin{eqnarray}
\bar M=\frac{r_+}{2}\left(1+\frac{r^2_+}{l^2}+\frac{2^{\gamma-1}q^{2\gamma}}{(4\gamma-3)r_+^{4\gamma-2}}\right).\label{M}
\end{eqnarray}
We can also obtain the Hawking temperature of the black hole from eq. (\ref{f}) as follows
\begin{eqnarray}
T=\frac{1}{4 \pi r_{+}}\left(1+8 \pi \bar P r_{+}^{2}-\frac{\left(2 q^{2}\right)^{\gamma}}{2 r_{+}^{(4 \gamma-2)}}\right).\label{T}
\end{eqnarray}
From eqs. (\ref{M}) and (\ref{T}) we will calculate the critical value of thermodynamical quantities which are presented in Sec. VI. Next we will give the modified first law of the four-dimensional EPYM AdS black hole thermodynamics in natural units ($\hbar=c=1$), i.e., the restricted phase space formulism.

\section{Joule-Thomson expansion}
\label{scheme3}
More recently, the authors of \cite{Okcu2017} have investigated the Joule-Thomson (JT) expansion for AdS charged black holes with the aim to confront the resulting features with those of Van der Waals fluids. The extension to the charged black hole solution in the presence of the quintessence field \cite{Ghaffarnejad2018} and rotating-AdS black hole \cite{Okcu2018} have also been considered. JT expansion \cite{Perry1934} is a convenient isoenthalpic tool that a thermal system exhibits with a thermal expansion. It is worth noting that when expanding a thermal system with a temperature $T$, the pressure always decreases yielding a negative sign to $\partial P$. In this section, we will investigate the Joule-Thomson expansion of the RN-dS spacetime. In JT expansion for the Van der Waals system as well as AdS black holes, gas/black hole phase is passed at high pressure through a porous plug or small value in the low-pressure section of an adiabatic tube, and the enthalpy remains constant during the expansion. The expansion is characterized by a change in temperature relative to pressure. The JT coefficient, which can describe the expansion process, is read as
\begin{eqnarray}
a=\left(\frac{\partial T}{\partial {P}}\right)_{H},
\end{eqnarray}
where the enthalphy is related with the internal energy
\begin{eqnarray}
H=U+ P V.
\end{eqnarray}
We can judge whether the system is in a cooling process or in a heating process by the positive or the negative for the JT coefficient. Namely, if the temperature of the system is increasing with the decreasing of pressure, the JT coefficient is negative and the system is in the heating process, while if the temperature is decreasing with the decreasing of pressure, the JT coefficient is positive and the system is in the cooling process.

In this part, we will investigate the JT expansion of the EPYM AdS black hole. As we know, when the system is in the JT expansion, the enthalpy of system is fixed. In the extended phase space, the mass parameter of AdS black hole is corresponding to the enthalpy and it only just keep a constant when the system is in a JT process. The Joule-Thomson coefficient can be expressed as
\begin{eqnarray}
H=\bar M,~~~~
a=\left(\frac{\partial T}{\partial \bar P}\right)_{H}
=\left(\frac{\partial T}{\partial \bar P}\right)_{\bar M,q}
=\left(\frac{\partial T}{\partial r_+}\right)_{\bar M,q}\bigg/\left(\frac{\partial \bar P}{\partial r_+}\right)_{\bar M,q}
\end{eqnarray}
In order to study the JT expansion of the system more easier, we rewritten the temperature and the pressure as
\begin{eqnarray}
T&=&\frac{1}{2\pi r_+}\left(-1+\frac{3\bar M}{r_+}-\frac{\gamma 2^{\gamma} q}{(4\gamma-3)r_+^{4\gamma-2}}\right)\label{Tjt},\\
\bar P&=&\frac{3}{8\pi r_+^2}\left(-1+\frac{2\bar M}{r_+}-\frac{2^{\gamma-1}q}{(4\gamma-3)r_+^{4\gamma-2}}\right)
\label{Pjt}.
\end{eqnarray}
From above equations, the JT coefficient becomes
\begin{eqnarray}
a&=&\frac{2r_+}{3}\frac{1-\frac{6\bar M}{r_+}+\frac{\gamma(4\gamma-1)2^\gamma q}{(4\gamma-3)r_+^{\gamma-2}}}{1-\frac{3\bar M}{r_+}+\frac{\gamma2^\gamma q}{(4\gamma-3)r_+^{\gamma-2}}}\nonumber\\
&=&\frac{4r_+}{3}\frac{2+8\pi\bar Pr_+^2-\frac{\left[2\gamma(4\gamma-1)-3\right]2^\gamma q}{2(4\gamma-3)r_+^{4\gamma-2}}}{1+8\pi \bar Pr_+^2-\frac{2^\gamma q}{2r_+^{4\gamma-2}}}.\label{ajt}
\end{eqnarray}
The JT coefficient will be divergent at the point $r_{+m}$, which is satisfied the following equation
\begin{eqnarray}
1-8\pi \bar Pr_{+m}^2-\frac{2^\gamma q}{2r_{+m}^{4\gamma-2}}=0.
\end{eqnarray}
It is very interesting that at the point $r_{+m}$ the hawking temperature in eq. (\ref{T}) is just zero, which indicates that the divergent point of the JT coefficient will be reveal the certain information of the extreme EPYM AdS black hole.
\begin{figure}[htp]
\includegraphics[width=0.4\textwidth]{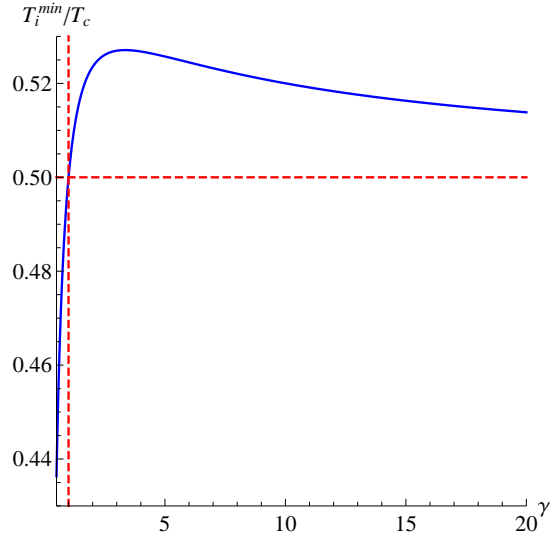}
\caption{The behavior of the ratio between the minimum inverse temperature and the critical one with the non-linear YM parameter.}\label{ratio}
\end{figure}

\begin{figure}[htp]
\centering
\subfigure[$~q=1$]{\includegraphics[width=0.4\textwidth]{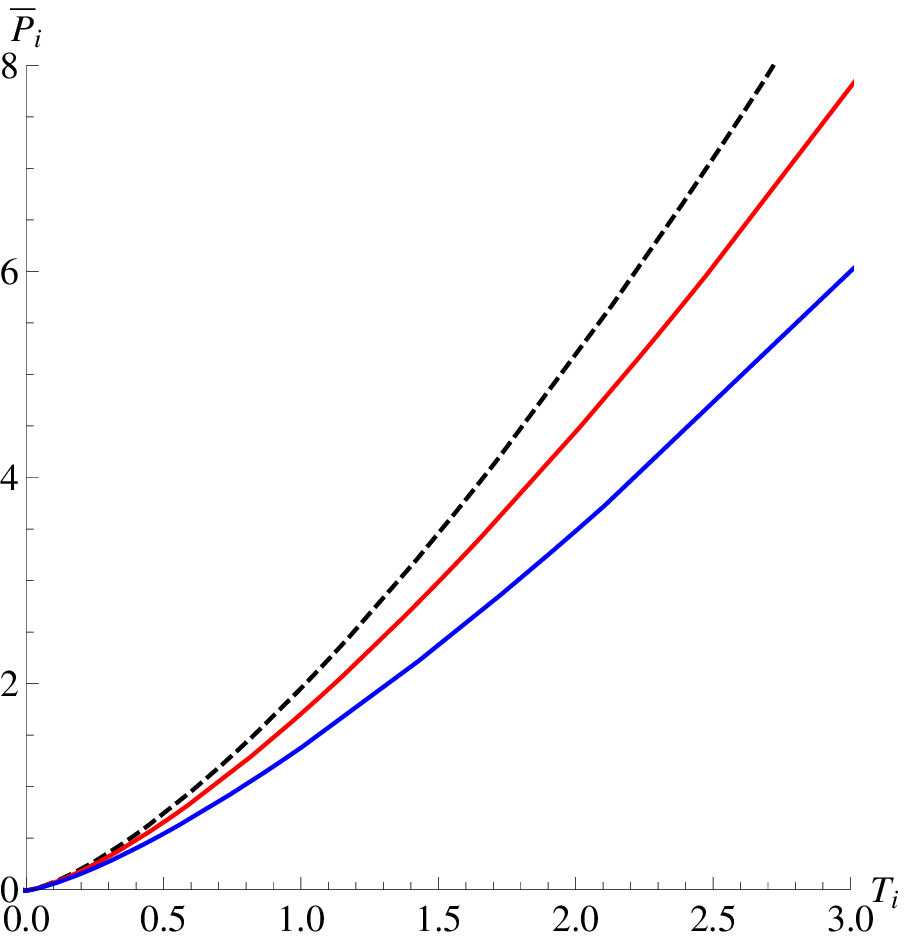}\label{Ti-Pi-gamma}}~~~~
\subfigure[$\gamma=0.85$]{\includegraphics[width=0.4\textwidth]{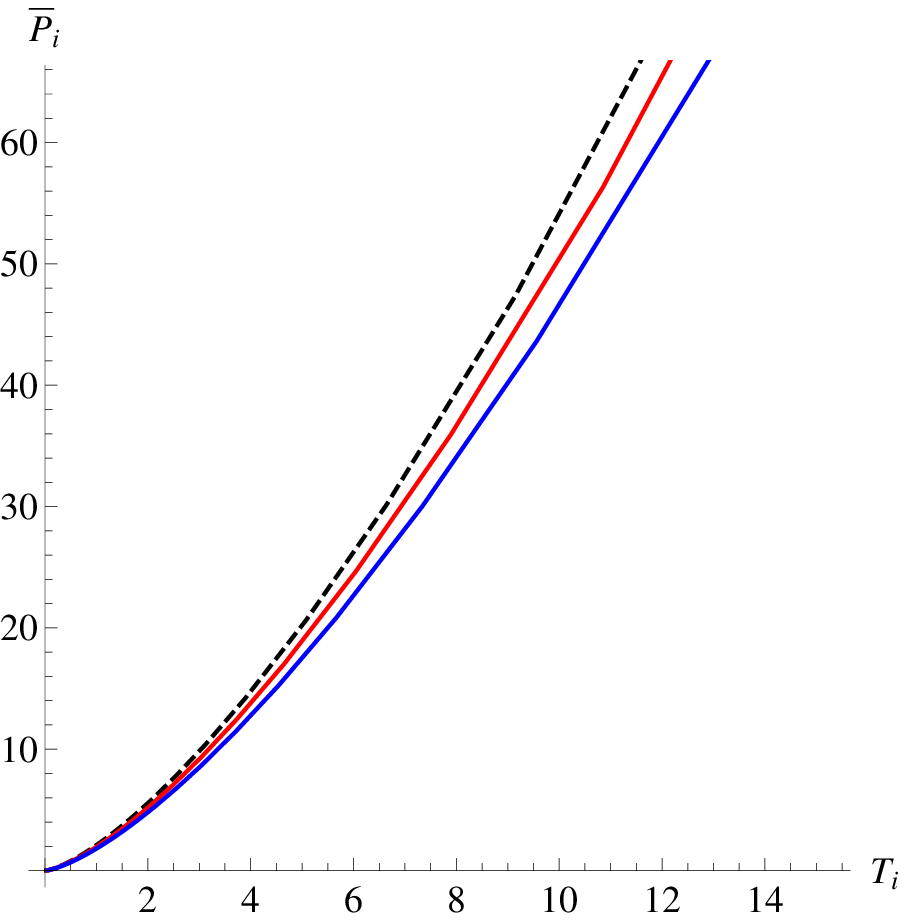}\label{Ti-Pi-q}}
\caption{The inverse curves $T_i-\bar P_i$ with different values of $\gamma$ (see the left) and of $q$ (see the right). In the left the non-linear charge parameter $\gamma$ is set to $0.85$ (the black dashed line), $0.9$ (the red thick line), and $1$ (the blue thick line). In the right the YM charge is set to $0.85$ (the black dashed line), $1$ (the red thick line), and $1.2$ (the blue thick line). }\label{Ti-Pi}
\end{figure}

In the following, we will focus on the minimum inverse temperature, the minimum inverse mass parameter, the isoenthalpic and inverse curves of this system. When the JT coefficient and the temperature are both zero in eq. (\ref{ajt}), the horizon radius satisfies
\begin{eqnarray}
r_i^{4\gamma-2}=\frac{\left[2\gamma(4\gamma-1)-3\right]2^{\gamma} q}{4(4\gamma-3)},\label{ri}
\end{eqnarray}
then substituting the above expression into eq. (\ref{Tjt}), the minimum inverse temperature reads
\begin{eqnarray}
T_i^{min}=\frac{8\gamma^2-10\gamma+3}{4\pi(8\gamma^2-2\gamma-3)}\left(\frac{\left[8\gamma^2-2\gamma-3\right]2^\gamma q}{4(4\gamma-3)}\right)^{-1/(4\gamma-2)}.
\end{eqnarray}
The ratio between the minimum inverse temperature and the critical one becomes
\begin{eqnarray}
\frac{T_i^{min}}{T_c}=\frac{\left(8\gamma^2-10\gamma+3\right)(4\gamma-1)}{4\left(8\gamma^2-2\gamma-3\right)(2\gamma-1)}
\left(\frac{8\gamma^2-2\gamma-3}{4\gamma(4\gamma-3)(4\gamma-1)}\right)^{-1/(4\gamma-2)}.
\end{eqnarray}
It is obviously that the ratio is independent with the YM charge and its behavior is exhibited in Fig. \ref{ratio}. Note that as $\gamma\rightarrow\infty$, the above ratio is approach to $1/2$.
\begin{figure}[htp]
\centering
\subfigure[$\gamma=0.85,~q=1$]{\includegraphics[width=0.4\textwidth]{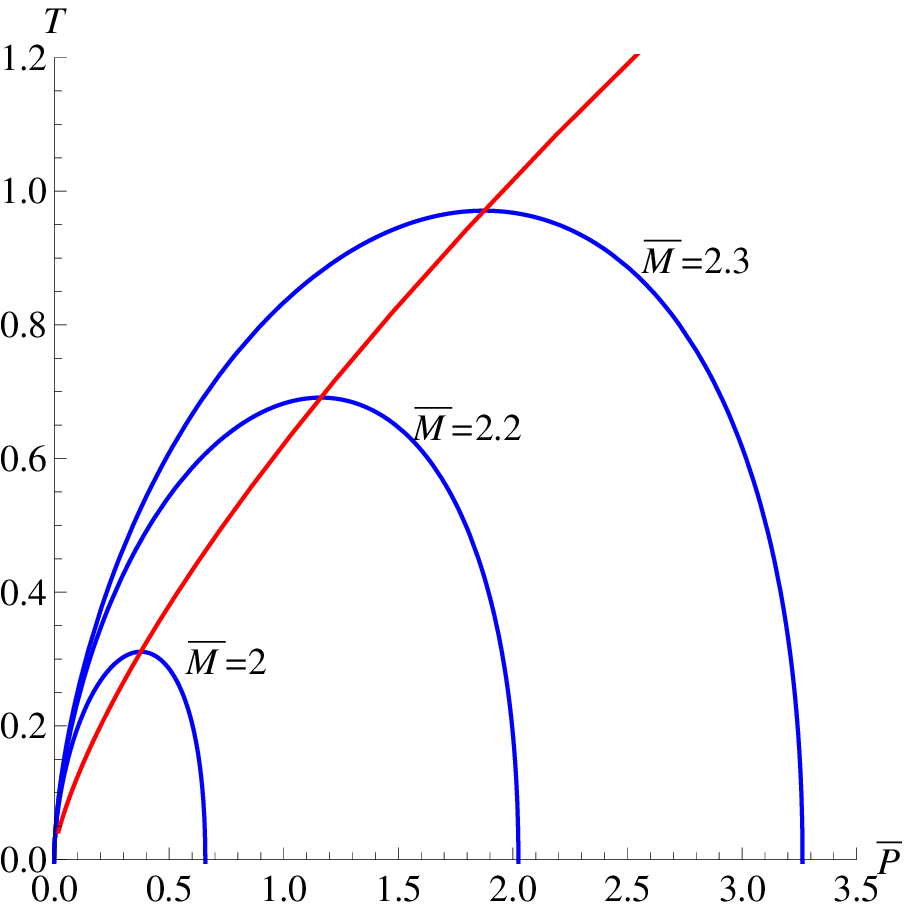}\label{T-P-851M}}~~~~
\subfigure[$\gamma=0.95,~q=1$]{\includegraphics[width=0.4\textwidth]{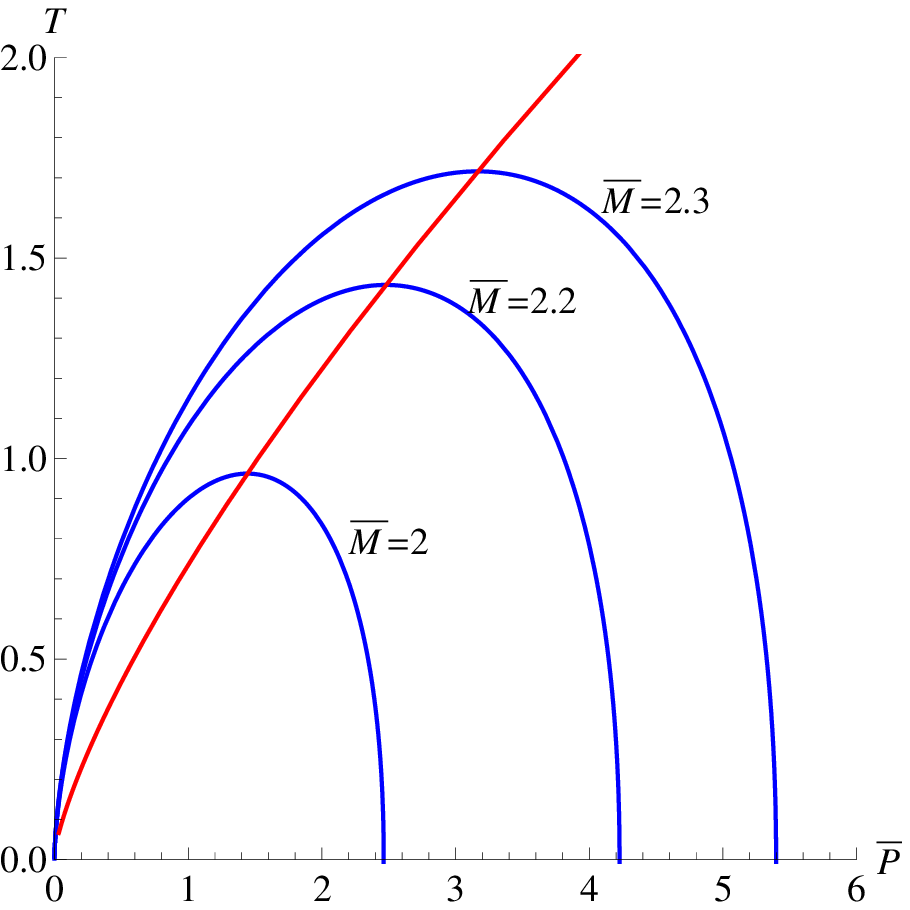}\label{T-P-951M}}\\
\subfigure[$\gamma=1.02,~q=1$]{\includegraphics[width=0.4\textwidth]{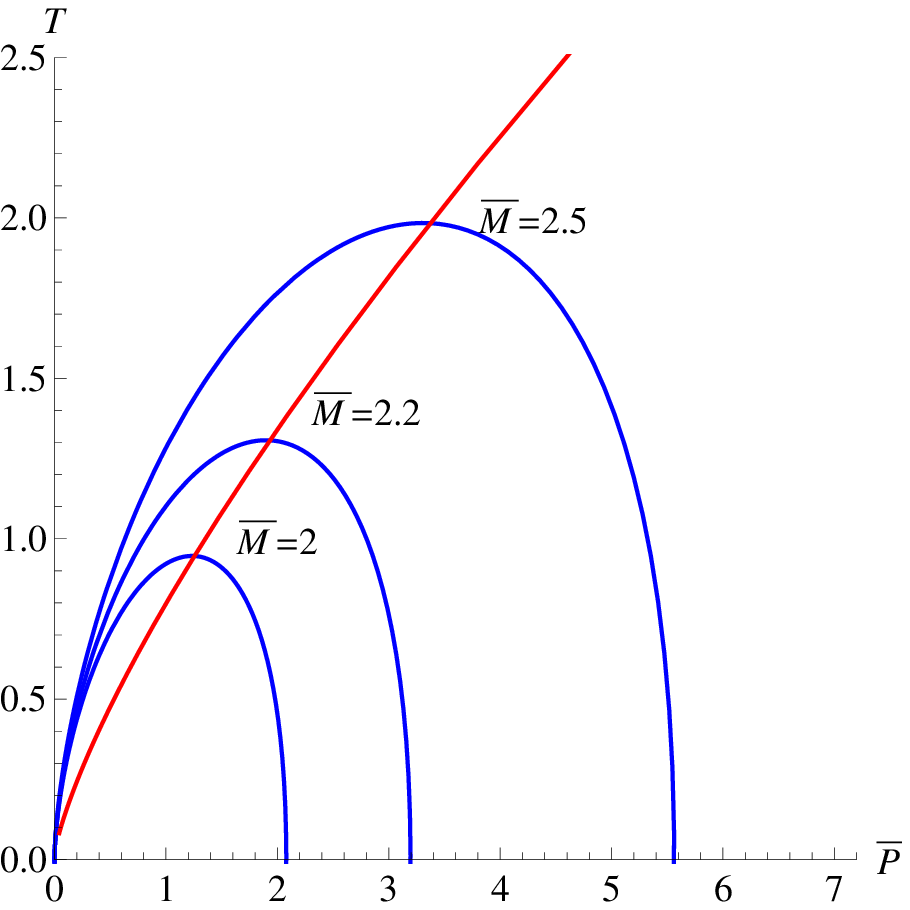}\label{T-P-1021M}}~~~~
\subfigure[$\gamma=0.95,~q=1.1$]{\includegraphics[width=0.4\textwidth]{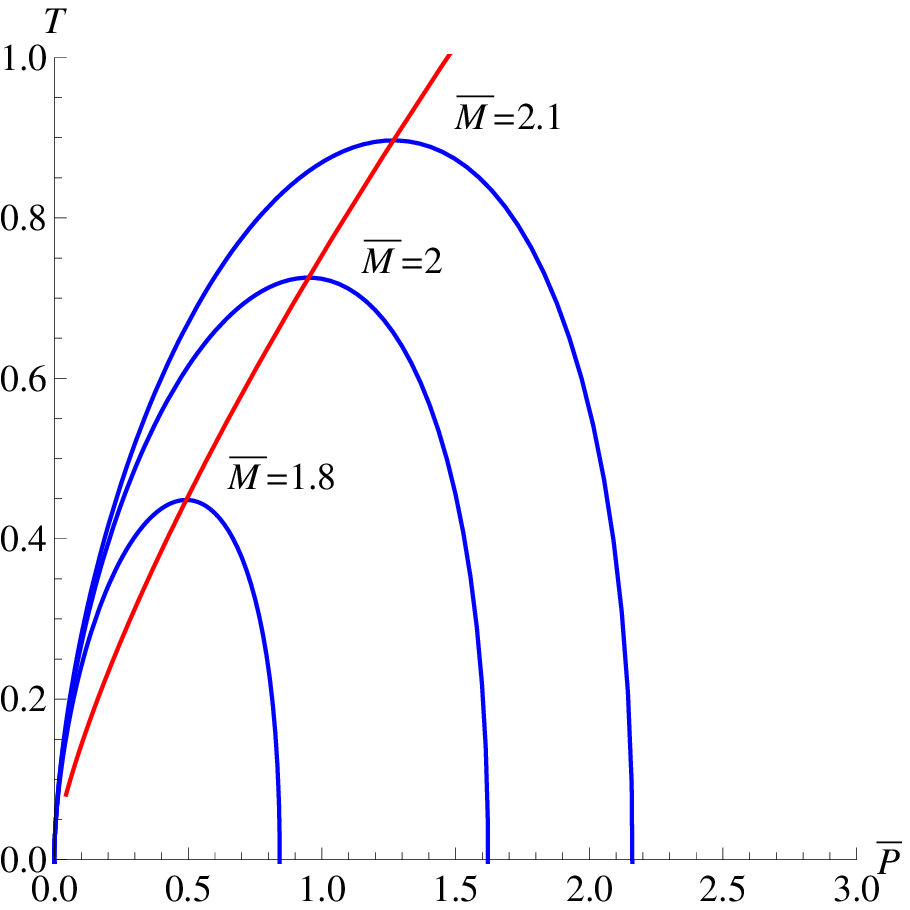}\label{T-P-9511M}}\\
\subfigure[$\gamma=0.95,~q=0.95$]{\includegraphics[width=0.4\textwidth]{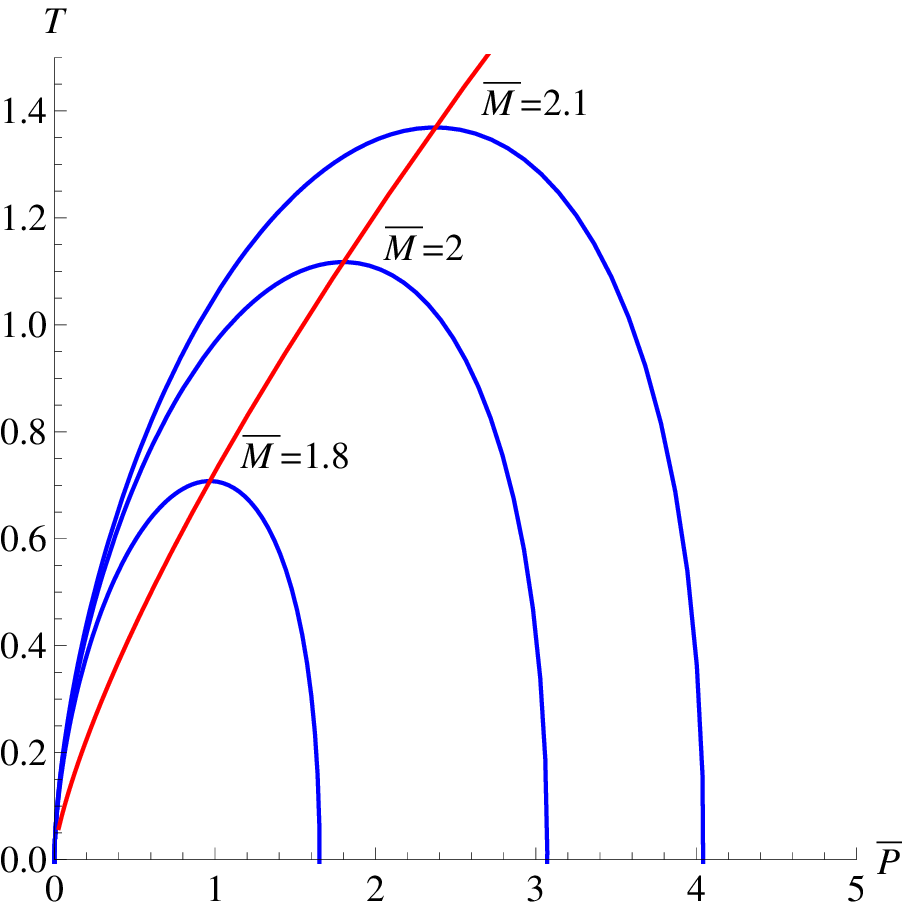}\label{T-P-9595M}}
\caption{The isoenthalpic and inverse curves with different values of the mass parameter.}\label{T-P-M}
\end{figure}
For the non-linear YM field (i.e., $\gamma\neq 1$), this ratio is not equal to $1/2$, which is just different from that for the linear YM field in this theory as well as the Einstein-Maxwell theory \cite{Okcu2018,Ahmed2018}. This difference is induced by the non-linear YM field, or maybe the modification of the thermodynamical volume. Especially, as $1<\gamma$ this ratio is bigger than $1/2$ and it is less than $1/2$ for $1/2<\gamma<1$. In addition when the pressure is zero and the temperature is the minimum inverse temperature, we can obtain the minimum inverse mass by substituting eq. (\ref{ri}) into eq. (\ref{M}) as
\begin{eqnarray}
\bar M_{min}=\frac{8\gamma^2-2\gamma-1}{2\left(8\gamma^2-2\gamma-3\right)}
\left(\frac{\left(8\gamma^2-2\gamma-3\right)2^\gamma q}{4(4\gamma-3)}\right)^{1/(4\gamma-2)}.
\end{eqnarray}
Since in the JT process of this system the black hole mass parameter $\bar M$ is unchanging, thus we can check out that whether the system is in a JT process through the minimum inverse mass parameter. That means a JT process of the system can be survive when $\bar M \geq\bar M_{min}$. Note that when $\gamma\rightarrow\infty$ the limitation of the minimum inverse mass approaches $\frac{1}{2^{3/4}}$ and it is independent with the YM charge. Furthermore as $3/4<\gamma$, $\bar M_{min}>0$ and it is decreasing with the non-linear YM charge parameter $\gamma$.

As the JT coefficient is zero, we can obtain the inverse pressure and temperature from eqs. (\ref{T}) and (\ref{ajt}) as
\begin{eqnarray}
\bar P_i&=&\frac{1}{8\pi r_i^2}\left(-2+\frac{\left[2\gamma(4\gamma-1)-3\right]2^\gamma q}{2(4\gamma-3)r_i^{4\gamma-2}}\right),\\
T_i&=&\frac{1}{4\pi r_i}\left(-1+\frac{\gamma2^\gamma q}{r_i^{4\gamma-2}}\right),
\end{eqnarray}
where the lower index ``$i$" stands for the inverse meaning. Therefore we can exhibit the inverse curve in the $\bar P-T$ plane with different values of the YM charge $q$ and the non-linear charge parameter $\gamma$ from above equations in Fig. \ref{Ti-Pi}. From Fig. \ref{Ti-Pi} we can see that there exists a inverse curve of the EPYM AdS black hole with the given parameters and it is not a circled one. The inverse temperature is increasing with the increasing of $q$ and $\gamma$. On the other hand, to better understand the Joule-Thomson expansion from eqs. (\ref{Tjt}) and (\ref{Pjt}) we can depict the isoenthalpic curves, the inverse curves, and the effects of $q$, $\gamma$ on them in Fig. \ref{T-P-M}. The result shows that the inverse curve divide the isoenthalpic one into two parts: one is the cooling phenomena with the positive slope of the $\bar P-T$ curves, the other is the heating process with the negative slope of the $\bar P-T$ curves. And both the inverse temperature and pressure are increasing with the increasing of the non-linear YM charge parameter, while they are decreasing with the YM charge.

\section{Discussions and Conclusions}
\label{scheme4}
In this manuscript we have analyzed the Joule-Thomson expansion of the EPYM AdS black hole in the expanded phase space. Considering the similar process of the gas expansion from a higher pressure section to a lower one by maintaining the fixed enthalpy, we applied it to the EPYM AdS black hole where the mass parameter is identified as enthalpy. Through the analysis of the Joule-Thomson coefficient we calculated the minimum inverse temperature and mass parameter, which could check out where a Joule-Thomson process of the system can be survive. We also presented the inverse curves in the $\bar P-T$ plane and the corresponding isenthalpic curves. Above the inverse curve we obtained the cooling region, while below the inverse curve it corresponds to the heating one. Especially the effect of the Yang Mills charge nonlinearity on the Joule-Thomson expansion was also investigated. The corresponding result can be summarized in the following
\begin{itemize}
\item{The ratio between the minimum inverse temperature and the critical temperature is independent with the YM charge and approaches to $1/2$ as $\gamma\rightarrow\infty$. When $\gamma=1$ it equals to $1/2$ and it is bigger than $1/2$ for $1<\gamma$, while less than $1/2$ for $1/2<\gamma<1$.}
\item{The minimum inverse mass parameter is independent with the YM charge as $\gamma\rightarrow\infty$, it is positive in the range of $3/4<\gamma$ and is decreasing with the increasing of the non-linear YM charge parameter. }
\item{Both the inverse temperature and pressure are increasing with the increasing of the non-linear YM charge parameter, while they are decreasing with the YM charge.}
\end{itemize}

\section*{Acknowledgements}

We would like to thank Prof. Ren Zhao for their indispensable discussions and comments. This work was supported by the National Natural Science Foundation of China (Grant No. 12075143), the science foundation of Shanxi datong university(2022Q1) and the teaching reform project of Shanxi datong university ( XJG2022234).


\begin{thebibliography}{99}

\bibitem{Bardeen1973}J. M. Bardeen, B. Carter and S. W. Hawking, {\it The Four laws of black hole mechanics}, Commun. Math. Phys. 31 (1973), 161-170.

\bibitem{Jacobson1995}T. Jacobson, {\it Thermodynamics of space-time: The Einstein equation of state,} Phys. Rev. Lett. 75 (1995), 1260-1263.

\bibitem{Padmanabhan2009}T. Padmanabhan, {\it Thermodynamical Aspects of Gravity: New insights,} Rept. Prog. Phys. 73 (2010), 046901.

\bibitem{Bekenstein1973a}J. D. Bekenstein, {\it Black holes and entropy,} Phys. Rev. D 7 (1973), 2333-2346.

\bibitem{Hawking1975}S. W. Hawking, {\it Particle Creation by Black Holes,} Commun. Math. Phys. 43 (1975), 199-22.

\bibitem{Hawking1983}S. W. Hawking and Don N. Page, {\it Thermodynamics of Black Holes in Anti-de Sitter Space}, Commun. Math. Phys. 87, 577-588 (1983).

\bibitem{Witten1998}E. Witten, {\it Anti-de Sitter space, thermal phase transition, and confinement in gauge theories}, Adv. Theor. Math. Phys. 2, 505 (1998), arXiv:hep-th/9803131.

\bibitem{Kastor2009}D. Kastor, S. Ray, and J. Traschen, {\it Enthalpy and the mechaniscs of AdS black holes}, Class. Quant. Grav. 26, 195011 (2009).

\bibitem{Maldacena1998} J. M. Maldacena, {\it The large N limit of superconformal field theories and supergravity}, Adv. Theor. Math. Phys. 2, 231 (1998).

\bibitem{Kubiznak2012} D. Kubiznak and R. B. Mann, {\it P-V criticality of charged AdS black holes}, JHEP 1207 (2012) 033, arXiv:1205.0559.

\bibitem{Cai2013}R.-G Cai, L.-M Cao, L. Li, and R.-Q Yang, {\it P-V criticality in the extended phase space of Gauss-Bonnet black holes in AdS space}, JHEP, 2013 (9), arXiv:1306.6233.

\bibitem{Wei2015}S.-W Wei and Y.-X Liu, {\it Insight into the microscopic structure of an AdS black hole from a thermodynammical phase transition}, Phys. Rev. Lett 115 (2015) 111302.

\bibitem{Altamirano2013}N. Altamirano, D. Kubiznak, and R. B. Mann, {\it Reentrant phase transitions in rotating anti-de Sitter black holes}, Phys. Rev. D 88 (2013) 101502, arXiv:1306.5756.

\bibitem{Frassin2014}A. M. Frassino, D. Kubiznak, R. B. Mann, and F. Simovic, {\it Multiple Reentrant Phase Transitions and Triple Points in Lovelock Thermodynamics}, JHEP, 09 (2014) 080, arXiv:1406.7015.


\bibitem{Dolan2014}B. P. Dolan, A. Kostouki, D. Kubiznak, and R. B. Mann, {\it Isolated critical point from Lovelock gravity}, Classical Quantum Gravity 31, 242001 (2014).

\bibitem{Wei2014}S.-W Wei and Y.-X Liu, {\it Triple points and phase diagrams in the extended phase space of charged Gauss-Bonnet black holes in AdS space,} Phys. Rev. D 90, 044057 (2014).

\bibitem{Li2022}M.-D Li, H.-M Wang, and S.-W Wei, {\it Triple points and novel phase transitions of dyonic AdS black holes with quasitopological electromagnetism}, Phys. Rev. D 105 (2022) 10.

\bibitem{Wei2020}S.-W Wei, Y.-X Liu, and R. B. Man, {\it Novel dual relation and constant in Hawking-Page phase transitions}, Phys. Rev. D 102 (2020) 10.

\bibitem{Johnsom2014}C. V. Johnson, {\it Holographic heat engines}, Class. Quant. Grav. 31 (2014) 205002.

\bibitem{Xu2017}H. Xu, Y. Sun, and L. Zhao, {\it Black hole thermodynamics and heat engines in conformal gravity}, Int. J. Mod. Phys. D 26 (2017) 13.
\bibitem{Kubiznak2017}D. Kubiznak, R. B. Mann, and M. Teo, {\it Black hole chemistry: Thermodynamics with Lambda}, Class. Quant. Grav. 34, 063001 (2017).


\bibitem{Okcu2017}O. Okcu and E. Aydiner, {\it Joule-Thomson expansion of the charged AdS black holes,} Eur. Phys. J. C 77 (2017) 24.

\bibitem{Okcu2018}O. Okcu and E. Aydiner, {\it Joule-Thomson expansion of Kerr-AdS black holes,} Eur. Phys. J. C 78 (2018) 123.

\bibitem{Mo2018} J.-X Mo, G.-Q Li, S.-Q Lan, and X.-B Xu, Phys. Rev. D 98 (2018) 124032.

\bibitem{Mo2020}J.-X Mo and G.-Q Li, Class. Quantum. Grav. 37 (2020) 045009.

\bibitem{Xing2021}J.-T Xing, Y Meng, and X.-M Kuang, Phys. Lett. B 820 (2021) 136604.

\bibitem{Liang2021}J. Liang, B. Mu, and P. Wang, Phys. Rev. D 104 (2021) 124003.

\bibitem{Ditta2022}A. Ditta, X. Tiecheng, G. Mustafa, et al, Eur. Phys. J. C 82 (2022) 1.

\bibitem{Johnson2018}C. V. Johnson, {\it Critical Black Holes in a Large Charge Limit}, Mod. Phys. Lett. A 33, 1850175 (2018), arXiv:1705.01154.

\bibitem{Johnson2018a}C. V. Johnson, {\it An Exact Model of the Power/Efficiency Trade-Off While Approaching the Carnot Limit}, Phys. Rev. D 98, 026008 (2018).
\bibitem{Mazharimousavi2008}S. H   Mazharimousavi and M. Halilsoy, {\it Black Hole solutions in Einstein-Maxwell-Yang-Mills-Gauss-Bonnet Theory}, J. Cosmol. Astropart. Phys. 12 (2008) 005.
\bibitem{Lorenci2002}V. A. De Lorenci, R. Klippert, M. Novello, and J. M. Salim, Phys. Rev. D 65, 063501 (2002).
\bibitem{Zhang2015a}M. Zhang, Z.-Y. Yang, D.-C. Zou, W. Xu, and R.-H. Yue, {\it P-V  criticality of AdS black hole in the Einstein-Maxwell-power-Yang-Mills gravity}, Gen. Rel. Grav. 47, 14 (2015), arXiv:1412.1197.
\bibitem{Mazharimousavi2007}S. H Mazharimousavi and M. Halilsoy, {\it 5D black hole solution in Einstein-Yang-Mills-Gauss-Bonnet thoery}, Phys. Rev. D 76, 087501 (2007).

\bibitem{Stetsko2021}M. M. Stetsko, {\it Static spherically symmetric black hole's solution in Einstein-Maxwell-Yang-Mills-dilaton theory}, International Journal of Modern Physics A 36, 05 (2021).

\bibitem{Du2021}Y.-Z Du, H.-F Li, F. Liu, R. Zhao, and L.-C Zhang, {\it Phase transition of non-linear charged Anti-de Sitter black holes}, Chinese Phys. C 45 (2021) 11.

\bibitem{Moumni2018}H. El Moumni, Phys. Lett. B 776, 124 (2018).

\bibitem{Bronnikov2001}K. A. Bronnikov, {\it Regular magnetic black holes and monopoles from nonlinear electrodynamics,} Phys. Rev. D 63 (2001), 044005, [arXiv:gr-qc/0006014].
\bibitem{Allahyari2020}A. Allahyari, M. Khodadi, S. Vagnozzi, and D. F. Mota, {\it Magnetically charged black holes from non-linear electrodynamics and the Event Horizon Telescope,} JCAP 2002 (2020), 003 [arXiv:arXiv:1912.08231].


\bibitem{Zhang2015}J.-L Zhang, R.-G Cai, and H. Yu, {\it Phase transition and thermodynamical geometry of Reissner-Nordstr\"{o}m-AdS black holes in extended phase space}, Phys. Rev. D 91, 044028 (2015).

\bibitem{Corda2011}C. Corda and H. J. Mosquera Cuesta, Astropart. Phys. 34, 587 (2011), arXiv:1011.4801 [physics.gen-ph].

\bibitem{Mazharimousavi2009}S. H. Mazharimousavi and M. Halilsoy, Phys. Lett. B 681, 190 (2009), arXiv:0908.0308.

\bibitem{Balakin2016}A. B. Balakin, J. P. S. Lemos, and A. E. Zayats, {\it Regular nonminimal magnetic black holes in spacetimes with a cosmological constant,} Phys. Rev. D 93, 024008 (2016); A. B. Balakin, H. Dehnen, and A. E. Zayats, {\it Nonminimal Einstein-Yang-Mills-Higgs theory: Associated, color, and color-acoustic metrics for the Wu-Yang monopole model}, Phys. Rev. D 76, 124011 (2007). A. B. Balakin and A. E. Zayats, {\it Non-minimal Wu-Yang monopole}, Phys. Lett. B 644, 294 (2007);S. H. Mazharimousavi and M. Halilsoy, {\it Einstein-Yang-Mills black hole solution in higher dimensions by the Wu-Yang ansatz}, Phys. Lett. B 659 (2008) 471.

\bibitem{Chakhchi2022}L. Chakhchi, H. El Moumni, and K. Masmar, {\it Shadows and optical appearance of a power-Yang-Mills black hole surrounded by different accretion disk profiles}, Phys. Rev. D 105 (2022) 064031.
\bibitem{Stetsko2020}M. M. Stetsko, {\it Static spherically symmetric Einstein-Yang-Mills-dilaton black hole and its thermodynamics}, Phys. Rev. D 101, 124017 (2020); M. M. Stetsko, {\it Static dilatonic black hole with nonlinear Maxwell and Yang-Mills fields of power-law type}, Gen. Relativ. Gravit. 53, 2 (2021).
\bibitem{Yerra2018}P. K. Yerra and C. Bhamidipati, {\it A Note on Critical Nonlinearly Charged Black Holes}, Mod. Phys. Lett. A 34 (2019) 27, arXiv:1806.08226.
\bibitem{Ghaffarnejad2018}H. Ghaffarnejad, E. Yaraie, and M. Farsam, {\it Quintessence Reissner Nordstr\"{o}m Anti de Sitter Black Holes and Joule Thomson effect,} Int. J. Theor. Phys, 2018.

\bibitem{Perry1934}J. H. Perry and C. V. Herrmann, {\it The Joule-Thomson effect of methane, nitrogen, and mixtures of these gases,} The Journal of Physical Chemistry, 39 (1934) 9.

\bibitem{Ahmed2018}C. Ahmed Rizwan, A. Naveena Kumara, D. Vaid and K. Ajith, {\it Joule-Thomson expansion in AdS black hole with a global monopole,} International Journal of Modern Physics A 33, 1850210 (2018)
%
%
%
%
%
%
%
%
%
%
%
%
%
%





\end{thebibliography}
\end{document}